\documentclass[showpacs,preprintnumbers,preprint,aps,amsmath,amssymb]{revtex4-1}
\usepackage{color}
\usepackage{graphics}
\usepackage{epsfig}
\usepackage{amsmath,amssymb}
\newcommand{\mathsym}[1]{{}}

\def\10{$SO(10)$}

\newcommand{\ba}{\begin{array}}
\newcommand{\ea}{\end{array}}
\newcommand{\be}{\begin{equation}}
\newcommand{\ee}{\end{equation}}
\newcommand{\beqa}{\begin{eqnarray}}
\newcommand{\eeqa}{\end{eqnarray}}
\def\321{$SU(3)\times SU(2)\times U(1)$}

\def\tbt{$t-b-\tau$ }

\def\sg{\tilde{g}}

\def\su{\tilde{u}}
\def\sd{\tilde{d}}
\def\st{\tilde{t}}
\def\sb{\tilde{b}}
\def\se{\tilde{e}}
\def\snu{\tilde{\nu}}
\def\stau{\tilde{\tau}}

\def\schi{\tilde{\chi}}

\begin{document}
\vspace*{1cm}
\title{Yukawa coupling unification in $SO(10)$ with positive $\mu$ and a heavier gluino}
\bigskip
\author{Anjan S. Joshipura}
\email{anjan@prl.res.in}
\author{Ketan M. Patel}
\email{kmpatel@prl.res.in}
\affiliation{Physical Research Laboratory, Navarangpura, Ahmedabad-380 009, India.
\vskip 1.0truecm}

\begin{abstract}
\vskip 0.5 truecm
The \tbt unification with positive Higgs mass parameter $\mu$ in the minimal supersymmetric standard
model prefers ``just so'' Higgs splitting and a light gluino $\lesssim$ 500 GeV which appears to be
ruled out by the recent LHC searches. We reanalyze constraints on soft supersymmetry breaking
parameters in this scenario allowing independent splittings among squarks and Higgs doublets at the
grand unification scale and show that it is possible to obtain \tbt unification and satisfy
experimental constraints on gluino mass without raising supersymmetry breaking scale to very high
value $\sim$ 20 TeV. We discuss the origin of independent squark and Higgs splittings in realistic
\10 models. Just so Higgs splitting can be induced without significantly affecting the \tbt
unification in \10 models containing Higgs fields transforming as $10+\overline{126}+126+210$. This
splitting arises in the presence of non-universal boundary conditions from mixing between $10$ and
other Higgs fields. Similarly, if additional matter fields are introduced then their mixing with the
matter multiplet $16$ is shown to generate the squark splitting required to raise the gluino mass
within the \tbt unified models with positive $\mu$.
\end{abstract}

\maketitle

\section{Introduction}
\label{intro}
Grand unified theories (GUTs) based on \10 group not only unify the gauge interactions but also
lead to a unified framework for the Yukawa couplings and hence fermion masses. In particular, \10
model with a 10-plet of higgs coupling dominantly to the third generation implies an equality
$y_t=y_b=y_\tau$ of the \tbt Yukawa couplings at the GUT scale. Quite independently, the
renormalization group (RG) running of the Yukawa couplings in a softly broken minimal supersymmetric
standard model (MSSM) can lead \cite{Ananthanarayan, Hall-Rattazzi, Carena, Pokorski, murayama,
sarid} to the \tbt unification at the GUT scale making the supersymmetric \10 broken to the MSSM at
the GUT scale an attractive theory of unification.

\tbt unification at the GUT scale is however not the most generic property of the MSSM but follows
only for a restricted set of boundary conditions for the soft supersymmetric breaking terms. These
restrictions mainly arise due to the need of significant threshold corrections \cite{Hall-Rattazzi,
Carena} to the $b$ quark mass required for the \tbt unification and difficulties in achieving the
radiative electroweak symmetry breaking (REWSB) in the presence of large $b$ and $\tau$ Yukawa
couplings \cite{Carena, Pokorski}. Both of these  depend on the soft breaking sector. It
is realized \cite{sarid, murayama} that \tbt unification generally requires departure from
the universal boundary conditions assumed within the minimal supergravity (mSUGRA) framework.
Universality of the gaugino masses is enforced by the \10 invariance if it is assumed that
supersymmetry (SUSY) is not broken at the GUT scale by a non-trivial representation contained in the
symmetric product of two adjoints of \10. In contrast, the soft masses $m_{16},~m_{10}$ for
sfermions belonging to $16_M$ and the Higgs scalars belonging to the $10_H$ representations are
allowed to be different and are also required to be so to obtain \tbt unification. In addition to
this \10 preserving non-universality, one also needs to introduce explicit \10 breaking
non-universality. Such non-universality can be induced spontaneously by a non-zero D-term (DT) which
introduces splitting within $16_M$ of squarks and $10_H$ of the Higgs simultaneously \cite{sarid,
Drees:1986vd}. In several situations, one also needs to assume that only the MSSM Higgs fields
$H_u,~H_d$ split at the GUT scale. This is termed as ``just so'' Higgs splitting (HS) \cite{raby}.

Restrictions placed on soft parameters by the \tbt unification, the LEP and LHC bounds on the masses
of the SUSY particles and other flavor violating observables have been worked out in detail in
number of papers \cite{raby, Baer:bc-for-pmu, Auto:2003ys, baer-dm, kraml, list}. Two viable
scenarios and their properties have been identified (see \cite{revtbt} for details and references
therein). These depend on the sign of the $\mu$ parameter of the MSSM. For example, one can achieve
an exact \tbt unification in mSUGRA itself for negative $\mu$. But this needs very heavy SUSY
spectrum with $m_{0} \sim 5-12$ TeV and $m_{1/2}\sim (1.5-2)m_{0}$ \cite{Auto:2003ys}. Also, perfect
\tbt unification with relatively light SUSY spectrum ($\sim$ 2 TeV) can be obtained with the
introduction of DT or purely Higgs splitting. This appears to be the best and testable scenario as
far as the \tbt unification is concerned. But the supersymmetric contribution to the muon $(g-2)$ is
negative in this case. This adds to the existing discrepancy between theory \cite{g2theory} and
experiments \cite{g2exp}. Scenario with positive $\mu$ proves better and allows the theoretical
prediction for $(g-2)$ to agree with experiments within 3$\sigma$. In this case, achieving \tbt
unification becomes considerably more difficult. The mSUGRA in this case at best allows unification
at 65\% level \cite{Auto:2003ys}. Even with non-universal boundary conditions, one needs specific
relations between the soft parameters, $m_{10}\sim 1.2 m_{16},~A_0\sim -2 m_{16} $ and $m_{1/2}\ll
m_{16}$ together with $\tan\beta\sim 50$ in order to achieve \tbt unification
\cite{Baer:bc-for-pmu}. The DT splitting in this case, allows unification at most 90\% level but
this requires $m_{16} \gtrsim 10$ TeV and a gluino mass $<500$ GeV. Just so HS works much
better than DT splitting and leads to the perfect Yukawa unification for $m_{16} \sim 10$ TeV but
gluino is still light. In both these scenarios, the sparticle mass spectrum is characterized
by lighter gluino which is within the reach of current LHC searches at $\sqrt{s}=7$ TeV, whereas all
scalar sparticles have masses beyond the TeV scale. The light gluino mainly decays through a three
body channel $\tilde{g}\rightarrow b\bar{b}\tilde{\chi}^0$ leading to multijets plus missing
transverse energy. The final states may also contain dileptons if $\st$ is lighter than $\sb$.
Recently, the ATLAS experiment with 2.05 fb$^{-1}$ data collected at $\sqrt{s}=7$ TeV has excluded
the light gluino masses below 620 GeV in \10+HS model \cite{ATLAS:2012ah}. As we will show later,
this experimental limit on gluino mass rules out \tbt unification better than 90\% for $m_{16} \sim
10$ TeV. The strong bound on the gluino mass follows from the need of suitable threshold corrections
in bottom quark mass to achieve \tbt unification which requires the hierarchy $m_{1/2}\ll m_{16}$.
One can thus raise the value of $m_{1/2}$ and hence the bound on the gluino mass by raising
$m_{16}$. The gluino mass can be pushed in this case beyond the present experimental limit but at
the cost of choosing $m_{16} \gtrsim 20$ TeV \cite{Baer-heavygluino}.

The best viable scenario with $\mu>0$ corresponds to just so Higgs splitting and large $m_{16}$ and
$m_{10}$. Theoretically, both these features are unsatisfactory. A large SUSY scale is unnatural and
just so HS breaks \10 explicitly. Just so HS can be indirectly introduced through the
right handed neutrinos at the intermediate scale. Apart from causing problem with the gauge coupling
unification, this case also does not do as well as the arbitrarily introduced just so HS,
see \cite{kraml}. We wish to discuss here possible ways to improve on both these aspects.
Specifically, we show that just so HS arises naturally in realistic \10 models containing additional
Higgs fields, e.g. the one transforming as $126,\overline{126},$ and $210$ representations
under \10. Realistic fermion masses can be obtained if these fields are present and the $SU(2)_L$
doublets contained in them mix with each other. Moreover, all the angles involved in such mixing
need not be small. We show that significant Higgs doublet mixing can generate just so HS in the
presence of  non-universal but \10 preserving boundary conditions at the GUT scale without
inducing any splitting between squarks or without significantly upsetting the \tbt Yukawa
unification.

\10 models also allow an interesting possibility of matter fields mixing among themselves 
\cite{dimo}. This leads to just so squark/slepton splitting similar to the just so HS occurring due
to Higgs mixing. Impact of such mixing was used earlier \cite{dimo} to obtain departure from the
$b$-$\tau$ unification that follows in $SU(5)$ or \10 models. Here we discuss phenomenological
implications of such mixing in the context of the \tbt unification. We discuss explicit example
leading to independent squark and Higgs splitting and show that the presence of such splittings
helps in raising the gluino mass prediction without raising the SUSY parameters $m_{16}$ to high
values around 20 TeV.

This paper is organized as follows. In the next section, we present a short review of the basic
features of the \tbt unification and discuss the existing phenomenological results. We also update
the existing results incorporating the recent limit on $B_s\rightarrow \mu^+\mu^-$ from LHCb
\cite{LHCb}. The viability of \tbt unified solutions in the presence of independent Higgs splitting
and squark splitting (SS) are discussed in Section \ref{heavier-gluino}. In Section \ref{HS-SS}, we
discuss how HS and SS can arise in the realistic versions of \10 models. The study is summarized
in the last section.

\section{\tbt unification in MSSM}
\label{tbtuni-MSSM}

In this section, we review numerical and analytic results presented in the literature
\cite{raby, Baer:bc-for-pmu, Auto:2003ys, baer-dm, kraml, list} in the context of the \tbt
unification. We have re derived several existing numerical results in a way which optimizes the rate
of the $B_s\rightarrow \mu^+\mu^-$ to make it consistent with the recent more stringent experimental
\cite{LHCb} bound without pushing the SUSY scale to a higher value. Before discussing this, we
summarize aspects of the \tbt unification which allows understanding of the salient key features.

\begin{figure}[h!]
 \centering
 \includegraphics[width=9cm,bb=0 0 400 267]{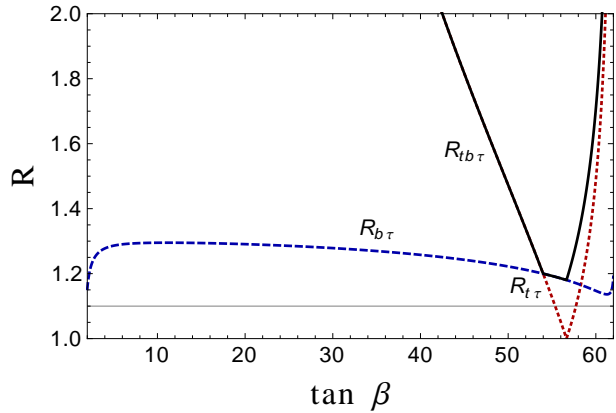}
 \caption{Tree level Yukawa unification as a function of $\tan\beta$. $R_{ij}$ defined in Eq.
(\ref{R}) measures the closeness of $y_i$ and $y_j$ at the GUT scale.}
 \label{fig-R}
\end{figure}

The hypothesis of the \tbt unification assumes that at the GUT scale
\be \label{uni}
y_b\equiv \frac{m_b}{v \cos\beta}=y_\tau\equiv \frac{m_\tau}{v\cos\beta}=y_t\equiv
\frac{m_t}{v\sin\beta}~, \ee
where $v\approx 174$ GeV. This equation is motivated by a simple \10 model containing only a single
10-plet Higgs. The appropriate choice of the free parameter $\tan\beta$ in Eq. (\ref{uni}) can
always allow equality of $y_t$ with $y_b$ or $y_\tau$ at the GUT scale. However it is well known
\cite{list} that $y_b$ and $y_\tau$ and hence all three of them derived from Eq. (\ref{uni}) using
the experimental values for fermion masses extrapolated to the GUT scale do not unify for any value
of $\tan\beta$. The degree of unification of three Yukawas or lack of it is usually measured by the
parameter
\be \label{R}
R_{tb\tau}\equiv R=\frac{{\rm Max.}(y_t,y_b,y_\tau)}{{\rm Min.}(y_t,y_b,y_\tau)}~.\ee
The parameter $R_{tb\tau}$ is defined at the GUT scale. Variation of $R_{tb\tau}$ with $\tan\beta$
obtained by using the tree level Yukawa couplings and fermion masses at $M_Z$ is shown in Fig.
(\ref{fig-R}).
For comparison we also show similar ratios $R_{b\tau}$ and $R_{t\tau}$ defined using only two of the
Yukawa couplings. The extrapolation from $M_Z$ to the GUT scale is done using the 1-loop RG
equations which are independent of the details of the soft SUSY breaking. Fig. (\ref{fig-R})
explicitly shows that the three Yukawas do not meet for any $\tan\beta$ and the reason is that $b$
and $\tau$ Yukawa couplings never meet. The best value of $R_{tb\tau}$ seen from Fig.
(\ref{fig-R}) is around 1.2. Thus any scheme which tries to achieve \tbt unification should do
better than this tree level value.

It is known that the tree level Yukawa couplings, particularly that of the $b$ quark receive
\cite{Hall-Rattazzi, Carena} significantly large radiative corrections once the supersymmetry
is broken. The corrected $y_b$ can be written as
\be \label{correctedyb}
y_b=y_b^{\rm tree} \cos\beta(1+\Delta y_b^{\tilde{g}}+\Delta y_b^{\tilde{\chi}}+...)~,\ee
where (...) contains the electroweak suppressed SUSY corrections, the standard model (SM)
electroweak corrections and logarithmic corrections which are sub dominant. The dominant correction
$\Delta y_b^{\tilde{g}}$ $(\Delta y_b^{\tilde{\chi}})$  induced by the gluino (chargino) exchange is
approximately given by
\cite{raby}
\be \label{threshold-gluino}
\Delta y_b^{\sg}\approx \frac{2 g_3^2}{3\pi} \mu \tan\beta
\frac{m_{\sg}}{m_{\sb_2}^2}~, \ee
\be \label{threshold-chargino}
\Delta y_b^{\schi^\pm}\approx \frac{y_t^2}{16 \pi^2} \mu \tan\beta
\frac{A_t}{m_{\st_2}^2}~,\ee
where $m_{\sb_2}$ ($m_{\st_2}$) is mass of the heaviest sbottom (stop).

The presence of $\tan\beta$ makes the radiative corrections significant. The corrections to
top Yukawa is inversely proportional to $\tan\beta$ while corrections to tau Yukawa is proportional
to $\tan\beta$ but electroweak suppressed. In order to achieve unification, one needs to reduce
$R_{b\tau}$ and hence $y_b$ compared to its tree level value by about 10-20\%. This requires that
$\Delta y_b^{\tilde{g}}+\Delta y_b^{\tilde{\chi}}$ in Eq. (\ref{correctedyb}) should be negative.
Since the gluino induced contribution dominates over most of the parameter space, one can make the
radiative corrections negative by choosing a negative $\mu$. As a result, models with negative
$\mu$ achieve \tbt unification more easily. For positive $\mu$, the chargino contribution has to
dominate over gluino and it should be negative. This can be satisfied with a negative $A_0$ and
light gluino with $|A_0|,~m_{16} \gg m_{1/2}$. As a result, all the scenarios of \tbt unification
with positive $\mu$ lead to a light gluino and very heavy SUSY spectrum as borne out by the detailed
numerical analysis \cite{raby, Baer:bc-for-pmu, Auto:2003ys, baer-dm, kraml}.

The requirement of a light gluino as argued above directly conflicts with the requirement of the
REWSB unless an explicit HS is introduced. This can be seen as follows. In large $\tan\beta$
limit, the REWSB can be achieved if
\beqa \label{rewsb}
-m_{H_u}^2&>& M_Z^2/2~, \nonumber\\
\Delta m_H^2&\equiv& m_{H_d}^2-m_{H_u}^2>M_Z^2~,\eeqa
where $m_{H_{u,d}}$ are soft scalar masses evaluated at the weak scale. Starting with a positive
value at $M_{GUT}$, $m_{H_u}^2$ gets driven to large negative values by a large $y_t$ and the first
equation gets satisfied. But the large $y_b,~y_\tau$ as required in the \tbt unification drives
$m_{H_d}^2$ even more negative and conflicts with the second requirement. In addition to the
Yukawas, the gaugino and scalar mass terms also contribute to the $\Delta m_H^2$. The former
contribution is positive while the latter is negative, see the semi analytic solution of the 1-loop
RG equations presented for example in \cite{murayama, Pokorski}.
\begin{figure}[h!]
 \centering
 \includegraphics[width=9cm,bb=0 0 400 267]{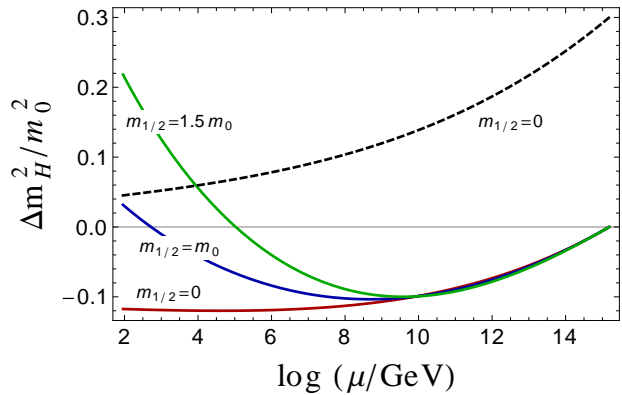}
 \caption{The solution of 1-loop RGE equation for $\Delta m_H^2 = m_{H_d}^2-m_{H_u}^2$ in mSUGRA for
$m_{0}(\equiv m_{10}=m_{16})=1$ TeV, $A_0=0$ and $\tan\beta$=50.}
 \label{fig-dm}
\end{figure}
In Fig. (\ref{fig-dm}), we show the running $\Delta m_H^2$ for different values of $m_{1/2}$. As can
be seen from the figure, second of Eq. (\ref{rewsb}) can be satisfied by choosing $m_{1/2}>m_{0}$
such that the gaugino induced contribution in $\Delta m_H^2$ dominates. On the other hand, small
$m_{1/2}$ and hence light gluino around $m_{\tilde{g}} \leq 500$ GeV is required if significant
corrections to $y_b$ is to be obtained in case of $\mu>0$. This corresponds to $m_{1/2} \leq 200$
GeV for which REWSB cannot be achieved unless one introduces splitting between Higgs fields at the
GUT scale itself. This is clearly seen from Fig. (\ref{fig-dm}). Moreover, the case in which only
Higgs splitting is considered is more favorable than the D-term splitting. This follows \cite{raby}
qualitatively from Eqs. (\ref{threshold-gluino}, \ref{threshold-chargino}) which implies
\be \label{ratio}
\frac{|\Delta y_b^{\sg}|}{|\Delta y_b^{\schi^\pm}|}\approx 11 \pi
\frac{m_{\sg}}{|A_t|}\frac{m_{\st_2}^2}{m_{\sb_2}^2}~. \ee
One finds $m_{\st_2}\sim m_{\sb_2}$ for mSUGRA as well for just so HS. The D-term splitting
introduces sfermion splitting together with Higgs splitting and leads to $m_{\st_2} > m_{\sb_2}$. As
a result, one needs to choose even a lighter gluino or a larger $|A_t|$ to suppress the gluino
induced corrections in $y_b$. As we will show later in this paper, the additional squark splitting
can instead reduce the ratio $m_{\st_2}/m_{\sb_2}$ and make it less than one. This allows
significantly higher gluino mass.

The above qualitative features are borne out by several numerical studies presented in a number of
papers \cite{Baer:bc-for-pmu, Auto:2003ys, baer-dm, kraml}. A list of different scenarios
proposed to achieve \tbt unification for $\mu>0$ is given in Table (1) in Ref. \cite{kraml}. Among
all the proposals, the \10 model with just so HS is found as the best scenario which leads to an
exact \tbt unification corresponding to $R=1$.  It is shown \cite{Auto:2003ys} that HS works
particularly well for large $m_{16}$ and the Yukawa unification $R\lesssim 1.02$ can be achieved if
$m_{16}\gtrsim10$ TeV. We update this analysis for the following reasons. In \cite{Auto:2003ys}, the
\tbt unified solutions were obtained considering the experimental constraint BF$(B_s\rightarrow
\mu^+\mu^-)^{\rm(exp)}<2.6\times10^{-6}$. The recent data collected by LHCb \cite{LHCb} experiment
has improved this bound significantly. The current limit BF$(B_s\rightarrow
\mu^+\mu^-)^{\rm(exp)}<4.5\times10^{-9}$ is three order of magnitude stronger than old bound. As a
result, all the solutions obtained in \cite{Auto:2003ys} are found inconsistent with new limit on
$B_s\rightarrow \mu^+\mu^-$ (see, Tabel (1) in \cite{Auto:2003ys}). We repeat the old analysis
considering the new limits on $B_s\rightarrow \mu^+\mu^-$ and $b\rightarrow s \gamma$. In addition,
we also consider the present constraints on $B\rightarrow \tau \nu_{\tau}$ \cite{Asner:2010qj} which
was not considered in the old analysis.

We use the ISASUGRA subroutine of ISAJET 7.82 \cite{isajet} in our numerical analysis. For given
boundary conditions (soft SUSY parameters at the GUT scale and $m_t$, $\tan\beta$ at the weak
scale), ISASUGRA solves full 2-loop MSSM RG equations and incorporates 1-loop SUSY threshold
corrections in all the MSSM sparticles and in the masses of third generation fermions. Moreover, it
checks for (a) non-techyonic solutions and (b) consistent REWSB using the minimization of one-loop
corrected effective MSSM Higgs potential. Once these conditions are satisfied, we calculate $R$
using Eq. (\ref{R}). Then using CERN's subroutine MINUIT, we minimize $R$. Finally, we calculate
branching factor for $b\rightarrow s \gamma$, $B_s\rightarrow \mu^+\mu^-$ using the IsaTools
package \cite{isajet} and $B\rightarrow \tau \nu_{\tau}$ using the expressions given in
\cite{Eriksson:2008cx}. For completeness, we also estimate the relic abundance of neutralino
dark matter $\Omega_{CDM}h^2$ and the SUSY contribution to anomalous magnetic moment of muon $\Delta
a_{\mu}$ (where $a_\mu=(g-2)/2$) using IsaTools. On the acquired solutions, we apply the following
constraints obtained from experimental data:
\beqa \label{expt-bounds}
&~ {\rm BF}(B_s \rightarrow \mu^+ \mu^-)~ & < \, 4.5 \times 10^{-9} ~~~~~~~~~~~~~~~
\text{\cite{LHCb}} \\
2.78 \times 10^{-4} \leq & {\rm BF}(b \rightarrow s \gamma) & \leq\, 4.32 \times 10^{-4} ~~
(3\sigma)~~~~~~\text{\cite{Asner:2010qj}} \\
0.62 \times 10^{-4} \leq &{\rm BF}(B \rightarrow \tau \nu_{\tau}) & \leq\, 2.66 \times 10^{-4}~~
(3\sigma)~~~~~~\text{\cite{Barlow:2011fu}}  \eeqa
Further, we impose the mass bounds given in PDG \cite{PDG} on all sparticles including the LEP
\cite{Schael:2006cr} bound on the mass of lightest Higgs ($m_{h} > 114.4$ GeV). We use
$m_t=172.9$ GeV in our analysis. The recent LHC limit on gluino mass is not considered here. We will
discuss it in detail in the next section.
\begin{table}[h!]
\begin{small}
\begin{math}
\begin{array}{|c|cccc|}
\hline
 \text{Parameter} & \text{Case I} & \text{Case II} & \text{Case III}& \text{Case IV} \\
\hline
 m_{16} 	& 10000 & 10000 & 15000 & 20000  \\
 m_{1/2} 	& 34.05 & 43.0 & 51.07 & 62.65  \\
 A_0/m_{16} 	& -2.29 & -2.26 & -2.29 &-2.45\\
 m_{10}/m_{16}  & 1.08 & 1.11 & 1.08 & 0.94\\
 \tan\beta	& 51.39 & 51.62 & 51.39 & 54.81\\
 \Delta m_H^2/m^2_{10}	& 0.25 &  0.28 & 0.25 & 0.39 \\
\hline
 R & {\bf 1.01} & {\bf 1.04} & {\bf 1.03} & {\bf 1.02}\\
\hline
 m_{\sg} & 345.0 & 367.9 & 487.7 & 634.9\\
 m_{\schi_{1,2}^0} & 49.6,~118.9 & 53.2,~127.1 & 73.6,~177.9 & 101.4,~241.7\\
 m_{\schi_{3,4}^0} & 6658.5,~6658.6 & 6460.2,~6460.3 & 9966.9,~9966.9 & 17102,~17103 \\
 m_{\schi_{1,2}^+} & 119.6,~6650.6 & 126.5,~6452.6 & 183.7,~9953.2 & 242.9,~17097 \\
 \hline
  m_{\su_{L,R}} & 9984.3,~9891.6 	~~&~~ 9988.4,~9874.7 ~~&~~ 14988,~14850~~&~~20010,~19782 \\
  m_{\sd_{L,R}} & 9984.7,~10043 	~~&~~ 9988.8,~10052 ~~&~~ 14988,~15075~~&~~20010,~20126 \\
  m_{\snu_{e,\tau}} & 9926.0,~7374.4 	~~&~~ 9912.4,~7274.3 ~~&~~ 14893,~11193~~&~~19841,~14636
\\
  m_{\se_{L,R}} & 9924.7,~10134 	~~&~~ 9911.2,~10159 ~~&~~ 14890,~15199~~&~~19838,~20312  \\
  m_{\st_{1,2}} & 2554.9,~3181.6 	~~&~~ 2506.2,~3110.1 ~~&~~
3957.5,~4859.7~~&~~5915.9,~6807.6 \\
  m_{\sb_{1,2}} & 2997.6,~3341.7 	~~&~~ 2902.4,~3226.2 ~~&~~
4724.0,~5097.9~~&~~6538.7,~6984.1 \\
  m_{\stau_{1,2}} & 3904.9,~7365.2 	~~&~~ 3664.9,~7278.2 ~~&~~ 6329.0,~11192~~&~~7549.7,~14617 
\\
\hline
 m_{h} & 126.0 & 125.8 & 127.8 & 125.0 \\
 m_{H} & 3463.6 & 4353.4 &  5664.2 & 8683.1\\
 m_{A} & 3441.4 & 4325.4 & 5627.4 & 8626.5\\
 m_{H^+} & 3465.5 & 4354.9 & 5665.3 & 8683.8 \\
\hline
 \text{BF}(b\rightarrow s\gamma) & 3.09\times10^{-4} & 3.07\times10^{-4} & 3.06\times10^{-4} &
3.06\times10^{-4}\\
 \text{BF}(B\rightarrow \tau \nu_{\tau})&
0.79\times10^{-4}&0.79\times10^{-4}&0.79\times10^{-4}&0.79\times10^{-4}\\
 \text{BF}(B_s\rightarrow \mu^+\mu^-)&  4.55\times10^{-9} & 4.22\times10^{-9} & 4.15\times10^{-9} 
& 3.96\times10^{-9}\\
\hline
 \Delta a_{\mu} & 0.024\times10^{-10} & 0.026\times10^{-10} & 0.008\times10^{-10}
& 0.008\times10^{-11}  \\
 \Omega_{CDM}h^2 & 2737 & 719 & 39866 &14115 \\
\hline
\end{array}
\end{math}
\end{small}
\vspace{0.0cm}
\caption{The benchmark solutions obtained for \tbt Yukawa unification in \10+HS model for positive
$\mu$. Different columns correspond to different cases discussed in the text. All masses are in GeV
units.}
\label{table:op}
\end{table}

The results of our numerical analysis are displayed in Table (\ref{table:op}). We study three
different cases:
\begin{enumerate}
 \item In case I, we do not impose constraint (\ref{expt-bounds}) and minimize $R$
for fixed $m_{16}=10$ TeV. The best unification found corresponding to $R=1.01$. The
solution predicts $\text{BF}(B_s\rightarrow \mu^+\mu^-)=4.55\times10^{-9}$ which is slightly above
the upper bound (\ref{expt-bounds}).
 \item The dominant contribution to $B_s\rightarrow \mu^+\mu^-$ in MSSM is proportional to
$m_A^{-4}$ \cite{Kane:2003wg} where $m_A$ is the mass of pseudo-scalar Higgs. Using this fact, we
simultaneously maximize $m_A$ and minimize $R$ in case II for $m_{16}=10$ TeV. As a result, we get a
lower value of $\text{BF}(B_s\rightarrow \mu^+\mu^-)$ which is consistent with experimental limit
(\ref{expt-bounds}). However one gets a slight declination in Yukawa unification in this case.
 \item Case III and IV correspond to an obvious way of decreasing SUSY contribution to flavor
violation namely, uplifting the SUSY scale. We take $m_{16}=15$ and 20 TeV which increase the masses
of all SUSY spectrum including $m_A$. The unification achieved in these cases is 97-98\%.
\end{enumerate}
It is clear from the results of our analysis that a very good \tbt Yukawa unification can still be
achieved with ``low'' $m_{16}$ in \10+HS model without violating the present experimental
constraint on $B_s\rightarrow \mu^+\mu^-$. The calculated values of $\text{BF}(b\rightarrow
s\gamma)$, $\text{BF}(B\rightarrow \tau \nu_{\tau})$ and $\Delta a_{\mu}$ shown in Table
(\ref{table:op}) are  almost similar to their standard model values. The SUSY contributions to
these processes are negligible due to the heavy sparticle spectrum one typically gets in \tbt
unified solutions for $\mu>0$. Among the other well known features of \tbt unification with positive
$\mu$ are relatively heavier Higgs $m_h \sim 125-130$ GeV arising  due to large $m_{16}$ and the
condition $A_0\sim -2 m_{16}$ \cite{Baer:bc-for-pmu, 125Higgs} and pure bino like lightest
neutralino which leads to the over abundance of the neutralino dark matter \cite{baer-dm}. One would
need additional mechanism, e.g. tiny $R$ parity violation \cite{rpv} to reduce this abundance.

\section{\tbt unification and heavier gluino}
\label{heavier-gluino}
As noted earlier, the \tbt unified solutions for positive $\mu$ generally require very light gluino
mass $\lesssim 500$ GeV. The direct SUSY searches at the LHC has now excluded $m_{\sg} \lesssim 620$
GeV in \10+HS model \cite{ATLAS:2012ah}. As a result, the solutions displayed in first three
columns in Table (\ref{table:op}) are ruled out. It is recently shown that consistent \tbt
unification with heavier gluino can be obtained by increasing $m_{16}$ \cite{Baer-heavygluino}. In
fact one needs $m_{16}\gtrsim 20$ TeV to evade the present LHC bound on gluino mass e.g. case IV in
Table (\ref{table:op}).  We propose here an alternate way to obtain heavier gluino in
\tbt unified solution without increasing $m_{16}$. As mentioned in Section \ref{tbtuni-MSSM}, the
ratio $|\Delta y_b^{\sg}|/|\Delta y_b^{\schi^\pm}|$ in Eq. (\ref{ratio}) can get an additional
suppression if $m_{\st_2}<m_{\sb_2}$. This allows \tbt unification with heavier gluinos. The
required mass hierarchy $m_{\st_2}<m_{\sb_2}$ can be obtained if the appropriate squark splitting is
introduced at the GUT scale. For example, consider $SU(5)$ invariant boundary conditions:
\beqa \label{sqarkBC}
m_{\tilde{Q}}^2&=&m_{\tilde{U}}^2=m_{\tilde{E}}^2\equiv m_{16}^2~, \nonumber \\
m_{\tilde{L}}^2&=&m_{\tilde{D}}^2 \equiv m_{16}^2+\Delta m_S^2~. \eeqa
The origin of such splitting in an \10 model will be discussed in the next section. As we will show
later in Eq. (\ref{bc2}, \ref{splitting2}), $\Delta m_S^2$ is allowed to take any values greater
than $-m_{16}^2$. The choice $\Delta m_S^2>0$ raises the mass of one eigenstate of sbottom squarks,
{\it i.e.} $m_{\sb_2}$, compared to its value obtained with universal squark masses at the GUT
scale. This leads to $m_{\st_2}<m_{\sb_2}$ at weak scale which is the case of our interest.

\begin{figure}[h!]
\centering
\includegraphics[width=8cm]{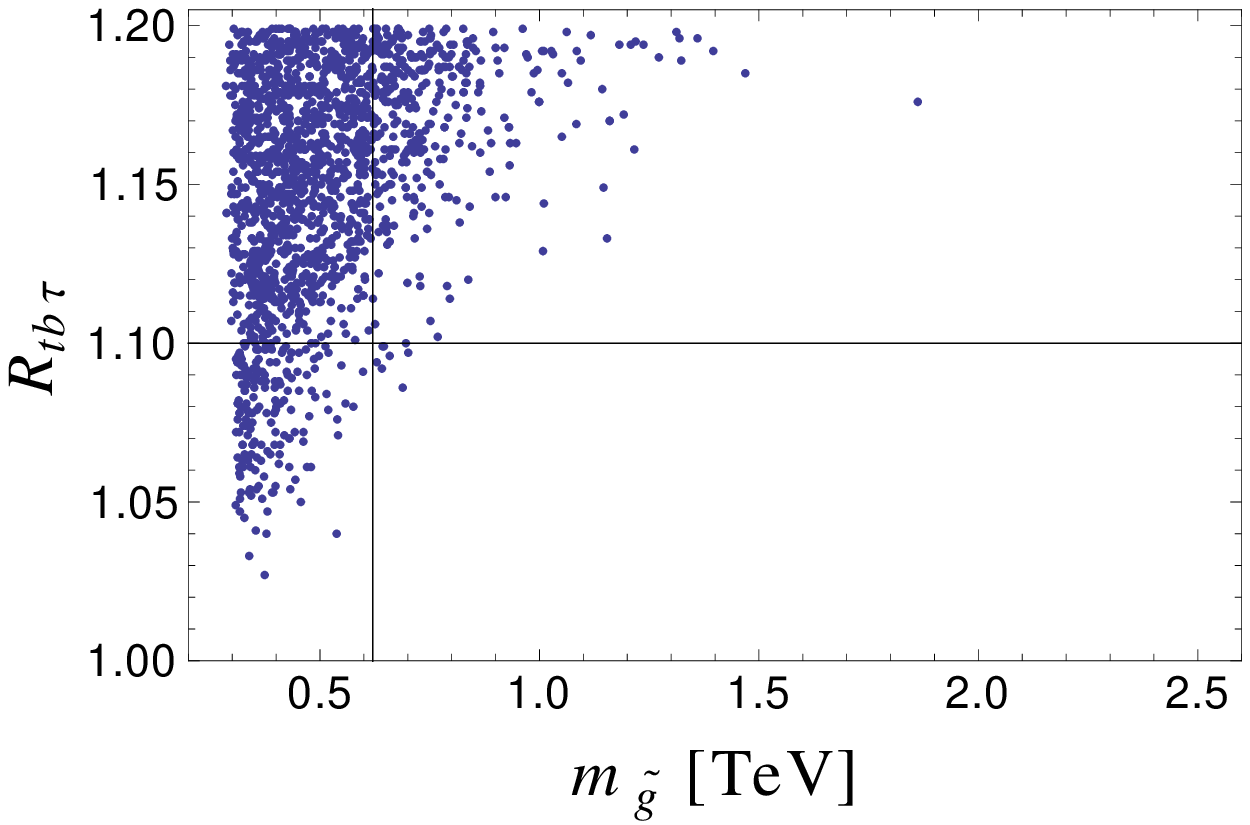}
\hspace{0.0cm}
\includegraphics[width=8cm]{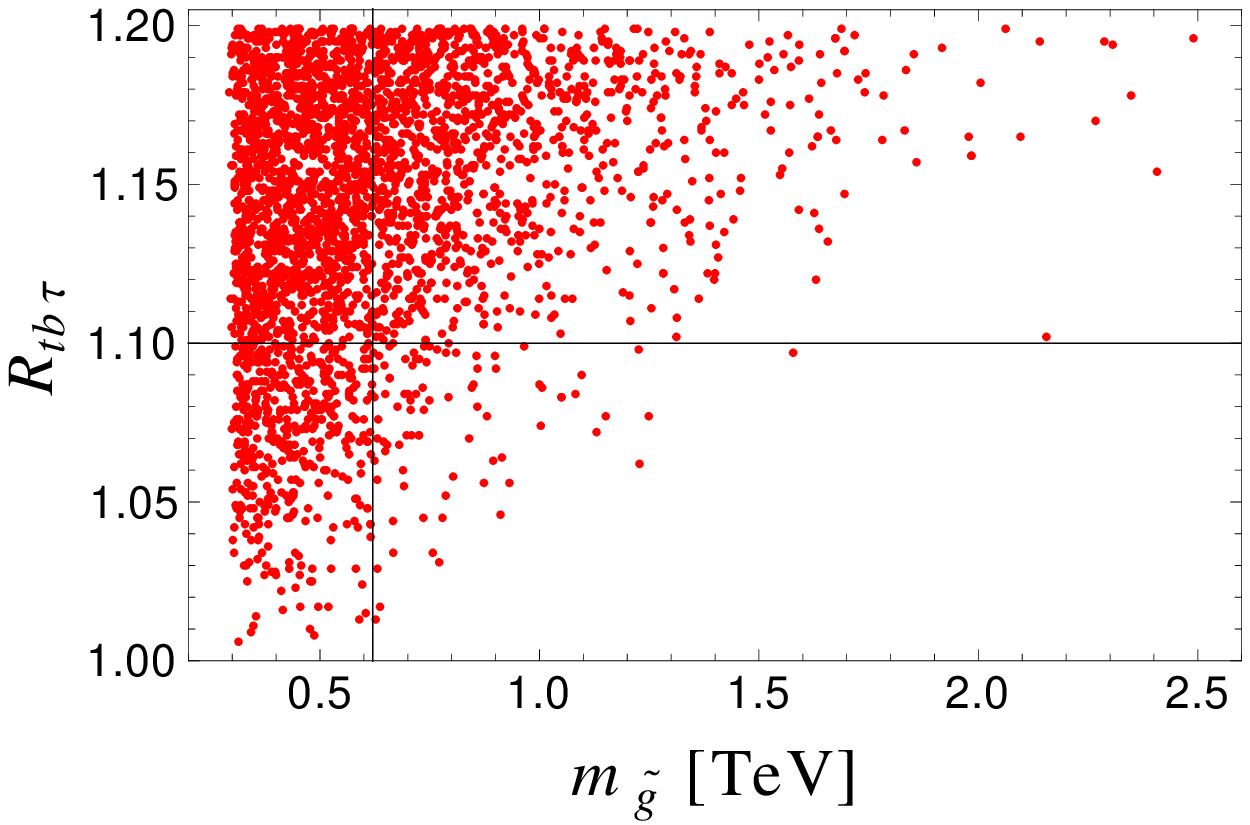}
\caption{Solutions of \tbt Yukawa unification as a function of gluino mass for $m_{16}=10$ TeV. The
figure in the left (right) panel shows the solutions obtained with only HS (HS+SS)
at the GUT scale in \10. The vertical line corresponds to the lower bound on gluino mass in
\10+HS model derived from the recent ATLAS data \cite{ATLAS:2012ah}. All the points shown are
consistent with various phenomenological constraints discussed in the text.}
\label{fig-gluino}
\end{figure}

To quantify the effect of squark splitting, we study the above case through detailed numerical
analysis. Fixing $m_{16}=10$ TeV we perform a random scan over all the remaining soft parameters and
$\tan\beta$. The analysis is performed for two scenarios: (1) only Higgs splitting, {\it i.e.}
for $\Delta m_S^2=0$ and (2) Higgs splitting + Squark splitting (HS+SS) with $\Delta m_H^2, \Delta
m_S^2>0$. We employ the same numerical technique and apply all the theoretical and experimental
constraints discussed in Section \ref{tbtuni-MSSM}. The results of numerical analysis are shown in
Fig. (\ref{fig-gluino}).

As can be seen from Fig. (\ref{fig-gluino}), the lower bound on gluino mass in \10+HS model rules
out the Yukawa unification $R<1.08$. In the presence of additional squark splitting, the unification
up to 99\% can be achieved without violating the present LHC limit on gluino mass. Also, a
relatively heavier gluino up to 1.5 TeV is obtained assuming at most 10\% deviation in Yukawa
unification without uplifting $m_{16}$. It can be also seen that the number of valid solutions
obtained with HS+SS are more compared to those obtained with only HS.
\begin{figure}[h!]
 \centering
 \includegraphics[width=9cm]{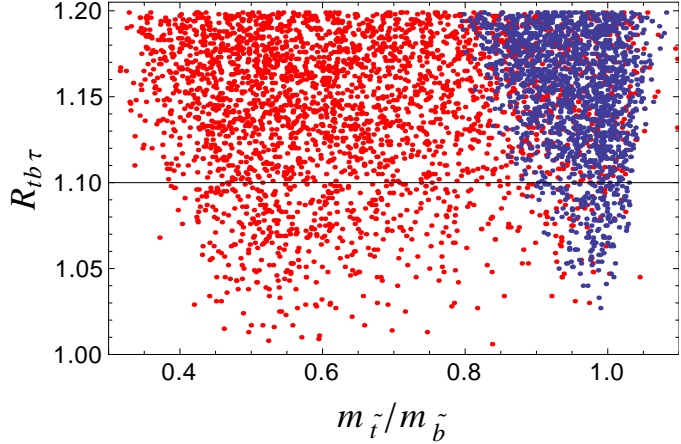}
 \caption{Squark mass ratio $m_{\st_2}/m_{\sb_2}$ obtained in HS (blue points) and HS+SS
(red points) models.}
 \label{fig-squark_ratio}
\end{figure}
In Fig. (\ref{fig-squark_ratio}), we show the ratio $m_{\st_2}/m_{\sb_2}$ in HS and HS+SS models.
The ratio substantially decreases in HS+SS model compared to its value without squark splitting.
As we mentioned earlier in this section, this allows a heavier gluino in the spectrum without
uplifting the SUSY braking scale. Note that with $m_{16}=10$ TeV, $m_{\sg}>1.5$ TeV cannot be
obtained for $R\lesssim 1.1$ even in the HS+SS model. This range of gluino mass is still in the
reach of LHC and its future operations at $\sqrt{s} = 14$ TeV can significantly constraint the
parameter space of HS+SS model if not rule it out completely.

\section{\tbt unification and realistic \10}
\label{HS-SS}
We consider two categories of \10 models. One in which Higgs sector is extended to obtain realistic
fermion masses and mixing and the other in which one introduces also additional matter multiplet at
$M_{GUT}$. The former class of models lead to just so HS and the latter also to an independent
squark splitting. We discuss them in turn.

\subsection{Just so HS in realistic \10}
\10 model containing a 10-plet of Higgs field $10_H$ allows the following term in the
superpotential:
\be \label{yukawa10}
Y_{10} 16_F 16_F 10_H~,\ee
where $16_F$ refers to the matter multiplet and $Y_{10}$ to the Yukawa coupling matrix in the
generation space.
The \tbt unification follows under two assumptions:\\
(A) Third generation of fermions obtain their masses only from Eq. (\ref{yukawa10}).\\
(B) The MSSM fields $H_u$ and $H_d$ reside solely in $10_H$.\\
One however needs to violate both these assumptions in order to obtain correct masses for all
fermions and non-trivial mixing among them. We estimate the effects of these violations on the
\tbt unification. We will take as an example a popular minimal renormalizable \10 model
\cite{babumohapatra} which is found to explain fermion masses and mixing in a number of situations,
for instance, see \cite{fermionmasses} and references therein.
The model contains a $\overline{126}_H$ field to generate neutrino mass and a $126_H$ to preserve
supersymmetry at the GUT scale. In addition, it has a Higgs transforming as $210_H$ representation
of \10 which breaks \10 to MSSM.

Due to the presence of additional Higgs fields particularly $210_H$, $SU(2)_L$ doublets residing in
various Higgs fields mix with each other. This mixing plays an important role in generating the
right type of the second generation masses \cite{babumohapatra}. But as we show below, this mixing
also generates just so Higgs splitting required to obtain \tbt unification if the soft masses of the
$10_H,~126_H$ and $210_H$ fields are non-universal. On the negative side, the presence of
$\overline{126}_H$ and the Higgs mixing also lead to departures from the exact \tbt unification.\\

The $10_H,~\overline{126}_H,~126_H,~210_H$ fields each contain four down-type and four up-type
standard model doublets. They mix and produce four mass eigenstates denoted  as $h_a^{u,d}$:
\be \label{mixing}
h_{b}^{u,d}=O_{ab}^{u,d}\phi_a^{u,d} ~,\ee
where $\phi_a^{u,d}$ are components of bi-doublets in $10_H,~\overline{126}_H,~126_H,~210_H$. One
assumes that through fine tuning only $H_u\equiv h_1^u$ and $H_d\equiv h_1^d$ remain light.
Consider now the soft mass terms of various \10 Higgs prior to the \10 breaking:
\be \label{soft}
V_{soft} \ni m_{16}^2 16_F^\dagger 16_F+m_{10}^2 10_H^\dagger 10_H + m_{126}^2(126_H^\dagger 126_H +
\overline{126}_H^\dagger
\overline{126}_H)+m_{210}^2 210_H^\dagger 210_H~.\ee
Here we have assumed equal masses for the $126_H$  and $\overline{126}_H$ fields to avoid non-zero
D-term. The masses of the other Higgs multiplets are taken non-universal. Substitution of Eq.
(\ref{mixing}) in Eq. (\ref{soft}) leads to
\beqa \label{boundry}
V_{soft}&\ni& m_{16}^2 (\tilde{Q}^\dagger \tilde{Q}+\tilde{U}^\dagger
\tilde{U}+\tilde{D}^\dagger \tilde{D}+\tilde{L}^\dagger \tilde{L}+\tilde{E}^\dagger \tilde{E})
\nonumber\\
&+& (m_{10}^2|O_{11}^u|^2+m_{126}^2(|O_{21}^u|^2+|O_{31}^u|^2)+m_{210}^2|O_{41}^u|^2)H_u^\dagger
H_u~\nonumber \\
&+&(u \rightarrow d).\eeqa
The above equation leads to the following boundary conditions at the GUT scale
\beqa \label{splitting}
m_{\tilde{Q}}^2&=&m_{\tilde{U}}^2=m_{\tilde{D}}^2=m_{\tilde{L}}^2=m_{\tilde{E}}^2=m_{16}^2,
\nonumber\\
m_{H_{u,d}}^2&=&m_{126}^2+|O_{11}^{u,d}|^2(m_{10}^2-m_{126}^2)+|O_{41}^{u,d}|^2(m_{210}^2-m_{126}
^2)~.
\eeqa
It is seen that the  Higgs mixing generated through Eq. (\ref{mixing}) has produced the desirable
splitting only among $H_{u,d}$ masses without splitting squarks from each other unlike in case of
the popular D-term splitting.

\begin{figure}[h!]
 \centering
 \includegraphics[width=9cm,bb=0 0 400 267]{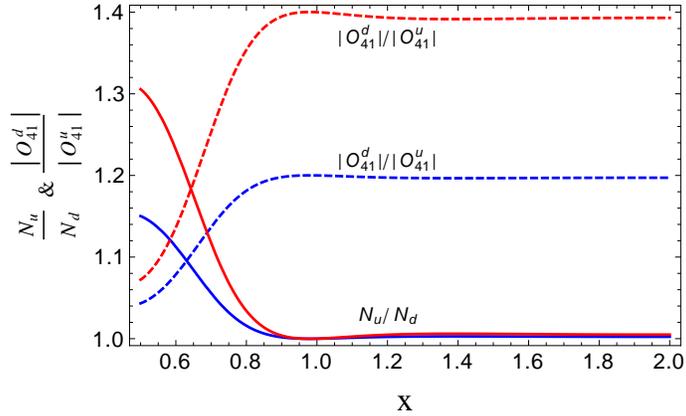}
 \caption{The Higgs mixing parameters $\frac{N_u}{N_d}$ (continuous lines) and
$|\frac{O_{41}^d}{O_{41}^u}|$ (dashed lines) as a function of $x$ in the minimal SUSY \10 model.
The blue (red) line corresponds to parameter $\epsilon=0.1$ (0.2).}
 \label{mixingplot}
\end{figure}

Let us now look at the impact of Higgs mixing on the \tbt unification. The presence of
$\overline{126}_H$ field modifies Eq. (\ref{yukawa10}) to
\be \label{yukawa}
Y_{10} 16_F 16_F 10_H+ Y_{126} 16_F 16_F \overline{126}_H,\ee
where $Y_{126}$ additional Yukawa coupling matrix. By substituting Eq. (\ref{mixing}) in Eq.
(\ref{yukawa}) one arrives at the charged fermion mass
matrices:
\beqa\label{yukawageneral}
M_d&=&\upsilon_d (Y_{10} O_{11}^d+Y_{126} O_{21}^d),\\
M_l&=&\upsilon_d (Y_{10} O_{11}^d-3Y_{126} O_{21}^d),\\
M_u&=&\upsilon_u (Y_{10} O_{11}^u+Y_{126} O_{21}^u),
\eeqa
where $\upsilon_{u,d}$ denote the vacuum expectation values of the neutral component of $H_{u,d}$.
We can go to a basis with $Y_{10}$ diagonal. Neglecting the contribution of $\overline{126}_H$ for
the time being, one has
\be\label{partialyb}
y_b=y_\tau=\frac{O_{11}^d}{O_{11}^u} y_t .\ee
Thus one source of the departure from the \tbt unification is the ratio
$O_{11}^d / O_{11}^u$. The parametric form of the matrices $O^{u,d}$ is worked out
\cite{charan, vissani} in the model under consideration and we closely follow the notation in
\cite{vissani}. This is based on the Higgs superpotential
\beqa \label{super}
W_H&=&M_{10}~10_H^2+M_{210}~210_H^2+M_{126}~126_H\overline{126}_H \nonumber \\
&+&\lambda~210_H^3+\eta~210_H 126_H \overline{126}_H+210_H
10_H(\alpha 126_H+\overline{\alpha}\overline{126}_H).\eeqa
The Higgs mass matrices and hence $O^{u,d}$ follow from the above superpotential after the \10
breaking and in the most general situation with \10 breaking to standard model one obtains (see,
Eqs.~(C18, C19) in \cite{vissani})
\beqa \label{Oud}
H_d=N_d\left( \frac{2
p_5}{x-1}\phi_{1}^d-\sqrt{6}\frac{\alpha}{\eta}(2x-1)(x+1)\frac{p_5}{p_3}\phi_2^d-\sqrt{6}\frac{\bar{
\alpha}}{\eta}(3x-1)(x^3+5 x-1)\phi_3^d+\bar{\alpha}\frac{\sigma}{m_{\phi }} p_3^\prime
\phi_4^d\right), \nonumber \\
H_u=N_u\left(\frac{2
p_5}{x-1}\phi_{1}^u-\sqrt{6}\frac{\bar{\alpha}}{\eta}(2x-1)(x+1)\frac{p_5}{p_3}\phi_3^u-\sqrt{6}\frac
{\alpha}{\eta}(3x-1)(x^3+5 x-1)\phi_2^u-\alpha\frac{\sigma}{m_{\phi}} p_3^\prime \phi_4^u\right),
\nonumber \\ \eeqa
where $N_{u,d}$ are overall normalization constants. $x$ is an arbitrary parameter and
$p_{3}, p_{3}^\prime, p_5$ are polynomial in $x$. The expressions of elements of $O^{u,d}$
can be read from the above equation. The ratio $O_{11}^u/O_{11}^d$ which measures deviation
from \tbt unification can be exactly or nearly one in a number of situations. Obvious case is the
limit $N_u = N_d \approx 1$ corresponding to the situation in which extra Higgs fields'
$\phi_{2,3,4}^{u,d}$ contribution to $H_{u,d}$ are sub-dominant. In general, various $\phi_a^u$ and
$\phi_a^d$ are components of $SU(2)_L\times SU(2)_R$ bi-doublets residing in
$10_H,~126_H,~\overline{126}_H,~210_H$ representations. They are thus distinguished by the $SU(2)_R$
group. One therefore automatically has $O^u=O^d$ as long as $SU(2)_R$ is unbroken. This happens
\cite{vissani} for $x=0$. One can then show from  Eq. (\ref{Oud}) that  $O^u=O^d$ in this
case. Another interesting limit corresponds to choosing $\alpha=\bar{\alpha}$ in
Eq. (\ref{Oud}). In this limit, $O^u\not= O^d$ but still $O_{11}^u=O_{11}^d$ thus one obtains exact
$y_b=y_t$ in Eq. (\ref{partialyb}) even in the situations where $SU(2)_R$ is broken. This case also
corresponds to $O_{41}^u=-O_{41}^d$ and thus HS also vanish in this limit as follows from
Eq. (\ref{splitting}). But mild deviation from the limit $\alpha=\bar{\alpha}$ can generate
sizable HS and approximate \tbt unification for large ranges in other parameter. This is illustrated
in Fig. (\ref{mixingplot}) where we show $\frac{N_u}{N_d}$ and $|\frac{O_{41}^d}{O_{41}^u}|$ as a
function of $x$ for two specific values of $\epsilon=1/2(\bar{\alpha}-\alpha)=0.1,~0.2$.
The remaining unknown parameters appearing in Eq. (\ref{Oud}) are equated to 1. It is seen that for
most values of the unknown parameter $x$ one obtains almost exact \tbt unification, {\it i.e.}
$\frac{N_d}{N_u}=1$ and non-zero HS \cite{footnote} as given in Eq. (\ref{splitting}).

Another threat to the \tbt unification comes due to the presence of the $Y_{126}$ Yukawa couplings
in Eq. (\ref{yukawageneral}). This effect is somewhat model-dependent and we estimate it by
specializing to the case of the second and third generations. We can write the charged fermion mass
matrices as
\beqa \label{massmatrices}
M_d&=&\upsilon_d\left(\ba{cc}
h_2+x_2&x\\
x&h_3+x_3\\ \ea \right);~~~
M_l=\upsilon_d\left(\ba{cc}
h_2-3x_2&-3x\\
-3x&h_3-3x_3\\ \ea \right);~\nonumber \\
M_u&=&\upsilon_u \frac{O_{11}^u}{O_{11}^d} \left(\ba{cc}
h_2+ s x_2& s x\\
s x&h_3+s x_3\\ \ea \right), \eeqa
where $s \equiv \dfrac{O_{21}^u}{O_{21}^d}\dfrac{O_{11}^d}{O_{11}^u}$. Here $h_{2,3}$ refer to
elements of the diagonal $Y_{10}O_{11}^d$ and $x_2,x,x_3$ to that of symmetric $O_{21}^d Y_{126}$.
Several of these can be determined from  the known masses and mixing. Approximate \tbt unification
is obtained with the hierarchy $x,x_2,x_3 \ll h_3$. Assuming $h_2 \ll x_2$ then leads to the
desirable mass relation $3 m_s=m_\mu$. Given this hierarchy one finds for real parameters
$$x_2\sim \frac{m_s}{m_b};~~~ x\sim m_b V_{cb}(1-s);~~~s\sim \frac{m_b}{m_s}\frac{m_c}{m_t}$$
The ratio $x_2/x_3$ remains undermined. If type II seesaw dominates and $Y_{126}$ is to
provide the neutrino masses and mixing then $x_2\sim x_3$ \cite{btau}. In this case, one finds
\beqa \label{uni2}
y_b&\approx&  h_3 \left(1+\frac{m_s}{m_b}\right)  \left(1+ O(V_{cb}^2)\right),\nonumber \\
y_\tau&\approx& h_3\left(1-3 \frac{m_s}{m_b}\right) \left(1+O(V_{cb}^2)\right),\nonumber\\
y_t&\approx& \frac{O_{11}^u}{O_{11}^d}h_3\left(1+\frac{m_c}{m_t}\right), \eeqa
where we have used Eq. (\ref{uni}). Using value for the mass ratios at the GUT scale for
$\tan\beta=50$ \cite{ross}, $\frac{m_s}{m_b}\approx 0.016, \frac{m_c}{m_t}\approx 0.0023$  already
implies about $6\%$ departure from  $b-\tau$ unification. The $m_s,m_c$ dependent terms in
Eq. (\ref{uni2}) get suppressed if one assumes $x_3\ll x_2$. In this case, one can still obtain
$b-\tau$ unification and reproduce the second generation masses and Cabibbo mixing. The Higgs mixing
factor can be nearly one as argued before. It is thus quite plausible that one can obtain almost
exact Yukawa unification not just in simplified but also in more realistic GUT based on \10.

\subsection{Squark splitting and \tbt unification}
The squark splitting can be induced by adding new matter fields having the same quantum numbers as
some of the squarks and letting them mix with the normal squarks. The minimal possibility at the \10
level is introduction of a $10_M$ field. This contains fields transforming under an $SU(5)$
subgroup of \10 as $5'_M+\overline{5}_M'$. Of these, the $\overline{5}^\prime_M$ can mix with 
$\overline{5}_M$ contained in the matter multiplet $16_M$ of \10. This mixing can generate the
squark splitting. Let us discuss details within a specific model which has been studied extensively
\cite{10m, malinsky} for several different reasons. The model contains three generations of $16_M$
and three copies of $10_M$. We will explicitly consider only the third generation and a $10_M$ in
the following. The Higgs sector consists of the usual $10_H$ supplemented by $16_H+\overline{16}_H$
and $45_H+54_H$. The $16_H$ is introduced to break the $B-L$ gauge symmetry and $45_H+54_H$ are
needed to complete the breakdown of \10 to SM. This model serves as a good example in which (a)
independent squark and Higgs splitting can be generated and (b) there exit ranges of parameters for
which \tbt unification is approximately maintained. We shall discuss only a part of the
superpotential relevant to describe Higgs and squark mixing, see \cite{malinsky}  for a general
discussion of the model. The superpotential of the model can be divided in two parts one describing
matter-Higgs interaction and the other describing Higgs-Higgs interactions:
\beqa \label{wm}
W_{M}&=&Y 16_M 16_M 10_H+F 16_M 10_M 16_H+M 10_M 10_M,\nonumber \\
W_H&=&M_{16} \overline{16}_H 16_H+M_{10} 10_H 10_H+H 16_H 16_H 10_H+H' \overline{16}_H
\overline{16}_H 10_H. \eeqa
The above superpotential is designed to respect the matter parity under which all the matter (Higgs)
fields are odd (even). This symmetry is essential for preventing renormalizable baryon and lepton
number violating terms. Scalar components of none of the matter fields acquire vacuum expectation
value (vev) and thus matter parity remains unbroken. Only fields appearing in the above
superpotential and acquiring the GUT scale vev are $1_H+\overline{1}_H$ contained in
$16_H+\overline{16}_H$ of \10. Thus after the GUT scale breaking, above  superpotial maintains
invariance under the  $SU(5)$ subgroup of \10. As a result, the mixing between the following $SU(5)$
components is allowed and arise from Eq. (\ref{wm}):
\be \label{mixed}
\left(\ba{c}
\overline{5}_{M}^l\\
\overline{5}_{M}^h \ea \right)=R(\theta)\left(\ba{c}
\overline{5}_{M}\\
\overline{5}_{M}^\prime \ea \right);~~~\left(\ba{c}
\overline{5}_{H}^l\\
\overline{5}_{H}^h \ea \right)=R(\gamma)\left(\ba{c}
\overline{5}_{H}\\
\overline{5}_{H}^\prime \ea \right);~~~\left(\ba{c}
5_{H}^l\\
5_{H}^h \ea \right)=R(\delta)\left(\ba{c}
5_{H}\\
5_{H}^\prime \ea \right), \ee
where $$R(j)=\left(\ba{cc}
\cos j & -\sin j \\
\sin j & \cos j \ea \right).$$
The fields with (without) prime are component of the original 10 ($16+\overline{16}$) of \10. It is
assumed that fields labeled with superscript $l$ are kept light by fine tuning and those with the
superscript $h$  pick up masses at the GUT scale. The mixing angles appearing above are related to
parameters in Eq. (\ref{wm}) and are explicitly given as 
\beqa \label{angles}
\tan\theta&=& \frac{F v_{16}}{M},\nonumber \\
\tan2\gamma&=& \frac{2 v_{16}(H M_{10}+H^\prime M_{16})}{(M_{16}^2-M_{10}^2)+
v_{16}^2(H^2-H^{\prime 2})},\nonumber \\
\tan2\delta&=& \frac{2 v_{16}(H^\prime  M_{10}+H
M_{16})}{(M_{16}^2-M_{10}^2)-v_{16}^2(H^2-H^{\prime 2})},\eeqa
where $v_{16}$ is a vev of $SU(5)$ singlets residing in $16_H+\overline{16}_H$. The light fields
transform under $SU(5)$ as $5_H+\overline{5}_H$ of Higgs and $10_M+1_M+\overline{5}_M$ which
together makes $16_M$ and $10_H$ of \10. The effective Yukawa couplings of these fields can be
worked out from Eqs. (\ref{wm}, \ref{mixed}) in a straightforward way. One finds
\be  \label{effective}
W_{eff}=Y s_\delta 10_M 10_M 5_{H}^l+(Y s_\gamma c_\theta+F s_\theta c_\gamma) 10_{M}
\overline{5}_{M}^l\overline{5}_{H}^l,\ee
where $s_j = \sin j$ and $c_j = \cos j$.

The Higgs mixing as given in Eq. (\ref{mixed}) leads to both the squark and Higgs splitting through
\10 invariant non-universal boundary conditions. Consider the following soft terms:
\beqa \label{vsoft}
V_{soft}&\ni& m_{16}^2 16_{M}^\dagger 16_M+m_{10}^2 10_M^\dagger 10_M+m_{16}^{\prime 2}
16_{H}^\dagger 16_H+m_{10}^{\prime 2} 10_H^\dagger 10_H\nonumber \\
&=& m_{16}^2(10_M^\dagger 10_M+\overline{5}_M^\dagger \overline{5}_M+1_M^\dagger
1_M)+m_{10}^2(5_M^{\prime \dagger}5_M^\prime + \overline{5}_M^{\prime
\dagger}\overline{5}_M^\prime)\nonumber\\
&+&m_{16}^{\prime 2}(10_H^\dagger 10_H+\overline{5}_H^\dagger \overline{5}_H+1_H^\dagger
1_H)+m_{10}^{\prime 2}(5_H^{\prime \dagger}5_H^\prime+\overline{5}_H^{\prime
\dagger}\overline{5}_H^\prime).\eeqa
Substitution of Eq. (\ref{mixed}) in Eq. (\ref{vsoft}) leads to the following boundary conditions
for the soft mass parameters of squarks and Higgs:
\beqa \label{bc2}
m_{\tilde{Q}}^2&=&m_{\tilde{U}}^2=m_{\tilde{E}}^2=m_{16}^2,\nonumber\\
m_{\tilde{L}}^2&=&m_{\tilde{D}}^2=m_{16}^2+s_{\theta}^2 (m_{10}^2-m_{16}^2),\nonumber \\
m_{H_d}^2&=&m_{16}^{\prime 2} c_{\gamma}^2+m_{10}^{\prime 2} s_{\gamma}^2,\nonumber \\
m_{H_u}^2&=&m_{16}^{\prime 2} c_{\delta}^2+m_{10}^{\prime 2} s_{\delta}^2 \eeqa
resulting in the following squarks and Higgs splitting
\beqa\label{splitting2}
\Delta m_S^2&\equiv& m_{\tilde{Q}}^2-m_{\tilde{D}}^2=s_{\theta}^2 (m_{10}^2-m_{16}^2),\nonumber \\
\Delta m_H^2&\equiv& m_{H_d}^2-m_{H_u}^2=(s_{\gamma}^2-s_\delta^2) (m_{10}^{\prime 2}-m_{16}^{\prime
2}).\eeqa
Thus model under consideration simultaneously generates independent mixing among squarks and
Higgses that result into squark splitting and Higgs splitting respectively. The mixing angles 
which generate these splitting also lead to departure from the exact \tbt unification as before. But
the exact $b$-$\tau$ unification follows for arbitrary values of these mixing angles. Even \tbt
unification also holds approximately in limiting cases, e.g. $F s_\theta\ll1$ and $H=H'$. Another
interesting limit corresponds to $s_\delta\approx s_\gamma c_\theta$ and $Fs_\theta\ll1$. In this
limit one gets $y_b=y_\tau\approx y_t$ and simultaneously non-zero splittings, $\Delta m_H^2\approx
s_\gamma^2 \Delta m_S^2$ for $m_{10}=m_{10}^\prime$ and $m_{16}=m_{16}^\prime$ in Eq.
(\ref{splitting2}). This relation automatically implies $\Delta m_H^2>0$ as required for the REWSB
when $\Delta m_S^2$ is chosen positive to raise the gluino mass limit.

\section{Summary}
\label{conclusion}
We have addressed two important issues in this paper in the context of the \tbt unification in \10
broken to MSSM  with a positive $\mu$ parameter. The existing analysis \cite{Baer:bc-for-pmu,
Auto:2003ys, baer-dm, kraml} have either assumed only  HS with degenerate squarks at the GUT
scale or a D-term splitting in which case  squark splitting is correlated to the HS. This scenario
appears to be ruled out save for very high SUSY scale $m_{16}$ around 20 TeV
\cite{Baer-heavygluino}. Detailed phenomenological analysis presented here shows that independent
and positive squark splitting $\Delta m_S^2$, Eq. (\ref{sqarkBC}) can change the allowed gluino mass
and scenario can be compatible with \tbt unification $R\sim 1.01$ and the recent ATLAS bound
$m_{\tilde{g}}>620$ GeV. Moreover, $R<1.1$ requires $m_{\tilde{g}}<1.5$ TeV. Thus viability of the
\tbt unification can be tested in future at LHC with $\sqrt{s}=14$ TeV. The other
issue addressed here concerns the origin of just so HS. While just so HS is introduced as a
phenomenological parameter in many works \cite{Baer:bc-for-pmu, Auto:2003ys, baer-dm, kraml}, its
origin is not justified in most of the existing analysis, see however \cite{halledm,sarid}. We have
taken here a concrete realistic model \cite{babumohapatra} used to understand fermion mass spectrum
and shown within it that just so HS is an automatic consequence  of the non-universal boundary
conditions and Higgs mixing. It is also shown that one can obtain sizable just so HS and
retain almost exact \tbt unification in this realistic model. Independent squark splitting required
to relax the gluino mass bound is also shown to follow in an extended model in which squarks mix
with additional matter multiplet. One may conclude based on the analysis presented here that \tbt
unification with positive $\mu$ is still phenomenologically viable and theoretically well-founded.\\

\begin{acknowledgments}
ASJ thanks the Department of Science and Technology, Government of India for support under the J.
C. Bose National Fellowship programme, grant no. SR/S2/JCB-31/2010. KMP would like to thank Partha
Konar for useful discussions on ISAJET package.\\
\end{acknowledgments}


\begin{thebibliography}{99}
\bibitem{footnote}
This remains true for most values of
$x$ which are not very near to zeros of the polynomial $p_3$ or $p_3^\prime$.
\end{thebibliography}
\end{document}